\documentclass[twocolumn,superscriptaddress,aps,pra]{revtex4}

\usepackage{amsfonts,amssymb,amsmath}
\usepackage[]{graphics,graphicx,epsfig}
\usepackage{amsthm}
\usepackage{color}

\def\identity{\leavevmode\hbox{\small1\kern-3.8pt\normalsize1}}

\newcommand{\ket}[1]{\left | #1 \right\rangle}
\newcommand{\bra}[1]{\left \langle #1 \right |}

\newcommand{\Tr}{\mathrm{Tr}}

\newcommand{\ketbra}[3]{\left| #1\rangle\langle#2\right|_{#3}}
\newcommand{\proj}[1]{\ket{#1}\bra{#1}}
\renewcommand{\epsilon}{\varepsilon}
\newcommand{\carr}{\circlearrowright}
\newcommand{\call}{\circlearrowleft}

\begin{document}
\title{Experimental distribution of entanglement with separable carriers}

\author{A.~Fedrizzi}
\email{fedrizzi@physics.uq.edu.au}
\affiliation{Centre for Engineered Quantum Systems \& Centre for Quantum Computer and Communication Technology,
School of Mathematics and Physics, University of Queensland, Brisbane, QLD 4072, Australia}
\author{M.~Zuppardo} 
\affiliation{School of Physical and Mathematical Sciences, Nanyang Technological University, Singapore}
\author{G.~G.~Gillett}
\affiliation{Centre for Engineered Quantum Systems \& Centre for Quantum Computer and Communication Technology,
School of Mathematics and Physics, University of Queensland, Brisbane, QLD 4072, Australia}
\author{M.~A.~Broome}
\affiliation{Centre for Engineered Quantum Systems \& Centre for Quantum Computer and Communication Technology,
School of Mathematics and Physics, University of Queensland, Brisbane, QLD 4072, Australia}
\author{M.~de~Almeida}
\affiliation{Centre for Engineered Quantum Systems \& Centre for Quantum Computer and Communication Technology,
School of Mathematics and Physics, University of Queensland, Brisbane, QLD 4072, Australia}
\author{M.~Paternostro}
\affiliation{Centre for Theoretical Atomic, Molecular, and Optical Physics, School of Mathematics and Physics, Queen's University Belfast, BT7 1NN, United Kingdom}
\author{A.~G.~White}
\affiliation{Centre for Engineered Quantum Systems \& Centre for Quantum Computer and Communication Technology,
School of Mathematics and Physics, University of Queensland, Brisbane, QLD 4072, Australia}
\author{T.~Paterek}
\email{tomasz.paterek@ntu.edu.sg}
\affiliation{School of Physical and Mathematical Sciences, Nanyang Technological University, Singapore}
\affiliation{Centre for Quantum Technologies, National University of Singapore, Singapore}

\begin{abstract}
The key requirement for quantum networking is the distribution of entanglement between nodes. Surprisingly, entanglement can be generated across a network without direct transfer---or \emph{communication}---of entanglement. In contrast to information gain, which cannot exceed the communicated information, the entanglement gain is bounded by the communicated quantum discord, a more general measure of quantum correlation that includes but is not limited to entanglement. Here, we experimentally entangle two communicating parties sharing three initially separable photonic qubits by exchange of a carrier photon that is unentangled with either party \emph{at all times}. We show that distributing entanglement with separable carriers is resilient to noise and in some cases becomes the only way of distributing entanglement through noisy environments.
\end{abstract}

\maketitle

Communication is the exchange of physical systems aimed at establishing correlations between communicating parties. The total correlations are quantified by the mutual information established between the sender and receiver~\cite{PhysRevA.72.032317}, and information theory states that the gain in mutual information cannot exceed the amount of \emph{communicated} information~\cite{Nature.461.1101,PhysRevLett.109.070501}:
\begin{equation}
\mathcal{I}_{\mathrm{final}} - \mathcal{I}_{\mathrm{initial}} \le \mathcal{I}_{\mathrm{comm}}.
\label{INFO_GAIN}
\end{equation}

This statement holds true both in classical and quantum physics, but it does \emph{not} generalise to quantum entanglement~\cite{RevModPhys.81.865}. Entanglement is a purely nonclassical type of correlation enabling tasks such as quantum teleportation~\cite{PhysRevLett.70.1895}, secure cryptography~\cite{PhysRevLett.67.661}, improved communication complexity~\cite{RevModPhys.82.665}, and quantum dense coding~\cite{PhysRevLett.69.2881}. 

Remarkably, Cubitt \emph{et al.} showed~\cite{PhysRevLett.91.037902} that quantum entanglement can be distributed between remote parties without communicating it: through the exchange of a carrier system that is never entangled with sender or receiver.
The gain in entanglement $\mathcal{E}$ between communicating sites is instead bounded from above \cite{PhysRevLett.109.070501,PhysRevLett.108.250501} by the amount of communicated  quantum discord $\mathcal{D}_{\mathrm{comm}}$
\begin{equation}
\mathcal{E}_{\mathrm{final}} - \mathcal{E}_{\mathrm{initial}} \le \mathcal{D}_{\mathrm{comm}}.
\label{ENT_GAIN}
\end{equation}
Quantum discord is a type of nonclassical correlation~\cite{JPhysA.34.6899,PhysRevLett.88.017901} which equates to entanglement in pure quantum states but can persist in mixed states with zero entanglement. Equation (\ref{ENT_GAIN}) implies that discord is a necessary resource for entanglement distribution with separable carriers. This provides an alternative method to conventional protocols which aim at entangling quantum nodes via transfer of preavailable entanglement~\cite{PhysRevLett.92.013602,PhysRevLett.92.197901}.

Here, we experimentally demonstrate entanglement distribution via separable carriers using polarisation-encoded single photons. We validate the discord bound in Eq.~(2) and show, both in theory and practice, that the implemented entanglement distribution protocol~\cite{PhysRevLett.109.080503} is robust against noise, despite distributing only a small amount of entanglement with each carrier. This is a crucial requirement for practical entanglement distribution between the nodes of a quantum network in a noisy environment. Moreover, we show the existence of a significant  range of parameters for which the use of our protocol based on the communication of separable carriers is the only way of establishing entanglement remotely.

\begin{figure}[!t]
\begin{center}
\includegraphics[width=\columnwidth]{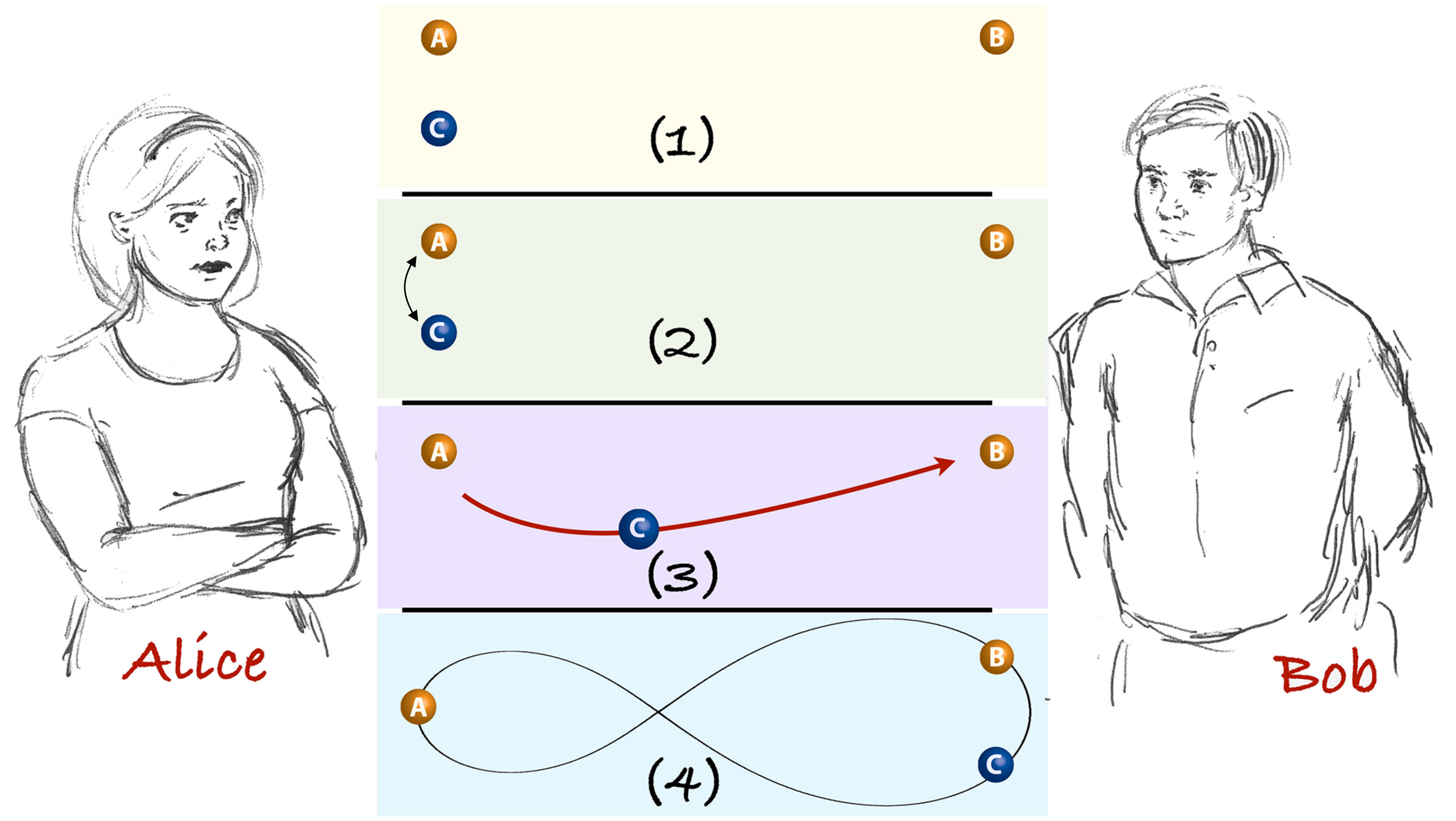}
\end{center}
\caption{(Quantum) communication scenario. Alice locally interacts her system $A$ with the carrier system $C$, which is then sent to Bob's site. It is possible to establish entanglement between their respective laboratories even though there was no initial entanglement between them and no entanglement is communicated. This is accomplished as follows: In step (1), the fully separable initial state of the three systems is prepared. In step (2), Alice applies a suitable operation on $A$ and $C$, which keeps the latter separable from the rest of the systems but creates entanglement between $A$ and joint systems made out of $B$ and $C$ together. In step (3), the unentangled carrier $C$ is transmitted to Bob. As shown in panel (4), this establishes entanglement between the laboratories of Alice and Bob.}
\label{FIG_COMM}
\end{figure}

A typical communication scenario is depicted in Fig.~\ref{FIG_COMM}.
Two parties, Alice and Bob, hold quantum systems $A$ and $B$, respectively. Alice uses a carrier quantum system $C$, which encodes her communication to Bob. We evaluate the entanglement gain between their respective laboratories, Eq.~(\ref{ENT_GAIN}), using  $\mathcal{E}_{\mathrm{final}} = \mathcal{E}_{A:CB}$, $\mathcal{E}_{\mathrm{initial}} = \mathcal{E}_{AC:B}$, and $\mathcal{D}_{\mathrm{comm}}{=}\mathcal{D}_{AB|C}$. Here, $\mathcal{E}_{X:Y}$ denotes the relative entropy of entanglement between $X$ and $Y$~\cite{PhysRevLett.78.2275}, while $\mathcal{D}_{X|Y}$ is the relative entropy of discord~\cite{PhysRevLett.104.080501}, also known as the one-way quantum deficit~\cite{PhysRevA.71.062307}.

In this protocol, the two-level systems $A$ and $B$ are prepared in a separable state $\alpha_{AB}$
that is a mixture of the four Bell states $\ket{\psi_\pm}_{AB}=\frac{1}{\sqrt{2}}(\ket{01} \pm \ket{10})_{AB}$ and $\ket{\phi_\pm}_{AB} = \frac{1}{\sqrt{2}}(\ket{00} \pm \ket{11})_{AB}$, with probability of occurrence $p_{\psi_\pm}$ and $p_{\phi_\pm}$, respectively.
Such a state is separable if and only if the highest probability in the mixture does not exceed $50\%$~\cite{PhysRevA.54.1838}.
The two-level carrier $C$ is initially with Alice and uncorrelated from the other systems, so that the overall initial state is taken as $\alpha = \alpha_{AB} \otimes \alpha_C$. Here $\alpha_C = \frac{1}{2}(\openone + c_x \sigma_x)$ with $\openone$ the identity matrix, $\sigma_k~(k=x,y,z)$ the Pauli $k$ matrix, and $c_x\in[-1,1]$.

Alice now generates the state $\beta = \mathcal{P}_{AC} \, \alpha \, \mathcal{P}_{AC}^\dagger$ by applying a controlled-phase gate ${\mathcal{P}}_{AC}=\ketbra{0}{0}{A}\otimes\openone_C+\ketbra{1}{1}{A}\otimes\sigma_{z,C}$ on her systems. We want the carrier qubit to remain separable from the other systems $\mathcal{E}_{AB:C}(\beta) = 0$, while system $A$ should become entangled with the subsystem composed of $B$ and $C$; i.e., we require $\mathcal{E}_{A:CB}(\beta) > 0$. Finally, carrier $C$ is transmitted to Bob and, as a result, the laboratories of Alice and Bob share entanglement.

We choose the initial state of $A$ and $B$ such that $C$ stays separable while maximising the entanglement in the $A|CB$ bipartition at the conclusion of the protocol. A possible instance is given by the $AB$-separable state
\begin{eqnarray}
\alpha_{AB} & = &  \frac{1}{4} \sum_{j=0}^1 \proj{z_j z_j} + \frac{1}{8} \sum_{j=0}^1 \proj{x_j x_j} \nonumber \\
& + & \frac{1}{8} \sum_{j=0}^1 \proj{y_j y_{1-j}},
\label{INITIAL_AB}
\end{eqnarray}
which is a mixture of two-qubit states formed by the eigenstates $\ket{k_j}$ of Pauli operators $\sigma_k$, with eigenvalue $(-1)^j$.
As a witness of entanglement, we use $\lambda^{\min}_{X|Y}$, the minimum eigenvalue of the partial transposition of the bipartite density matrix $\beta_{X|Y}$ with respect to sub-system $X$~\cite{PhysRevLett.77.1413}. Since the theoretical states considered and experimental states measured yield at most one negative eigenvalue, this witness is related to the negativity $\mathcal{N}$ ~\cite{PhysRevA.65.032314}---an entanglement measure---by $\mathcal{N}_{X|Y}{=}(\vert \lambda^{\min}_{X|Y}\vert-{\lambda^{\min}_{X|Y})/2}$.

Within the class of initial states $\alpha$ on which one applies ${\cal P}_{AC}$, the state composed of Eq.~\eqref{INITIAL_AB} and $\alpha_C$ with $c_x = -\frac{1}{2}$ gives $\mathcal{N}_{A|BC} = 1/16 =  0.0625$, the highest possible amount of entanglement that can be distributed via separable states \cite{PhysRevLett.109.080503}. We focus on ${\cal N}$ as its presence in $\beta$ guarantees that (i) the entanglement established between Alice and Bob can be localised into entanglement between $A$ and $B$ using only local operations performed at Bob's site~\cite{PhysRevLett.109.070501} and
(ii) such localised entanglement is distillable~\cite{PhysRevLett.78.574}.
Therefore, by repeating this protocol a sufficient number of times and performing entanglement distillation, 
one can in principle obtain maximally entangled pairs between Alice and Bob without ever communicating entanglement between them.

\begin{figure}[!b]
\begin{center}
\includegraphics[width=\columnwidth]{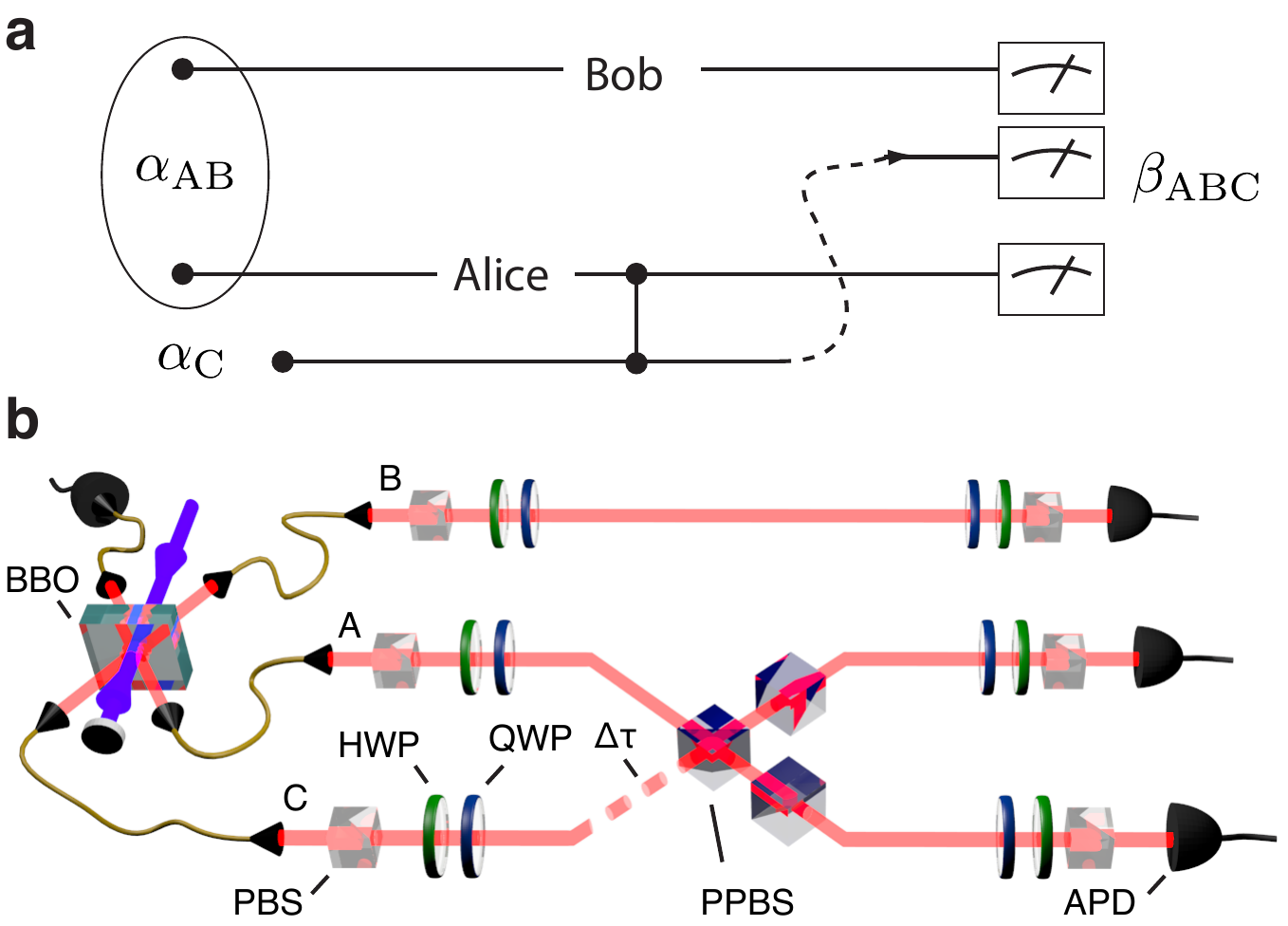}
\end{center}
\caption{Entanglement distribution scheme. (a) Equivalent quantum circuit diagram for our protocol. (b) Two pairs of single photons at $820$ nm are created via spontaneous parametric down-conversion in a $\beta$-barium borate crystal (BBO) pumped by a frequency-doubled femtosecond Ti:Sapphire laser. One photon serves as a trigger, while the other three are initialised with polarising beamsplitters (PBS) and half-wave (HWP) and quarter-wave plates (QWP). The photons representing systems $A$ and $C$ are subjected to a probabilistic controlled-phase gate based on two-photon interference at a partially polarising beamsplitter (PPBS) \cite{PhysRevLett.95.21}. Projective measurements are carried out with a combination of HWPs, QWPs and PBSs, before the photons are detected by single-photon avalanche photodiodes (APD) connected to a coincidence logic.}
 \label{FIG_SETUP}
\end{figure}

\begin{figure}[!b]
\begin{center}
\includegraphics[width=\columnwidth]{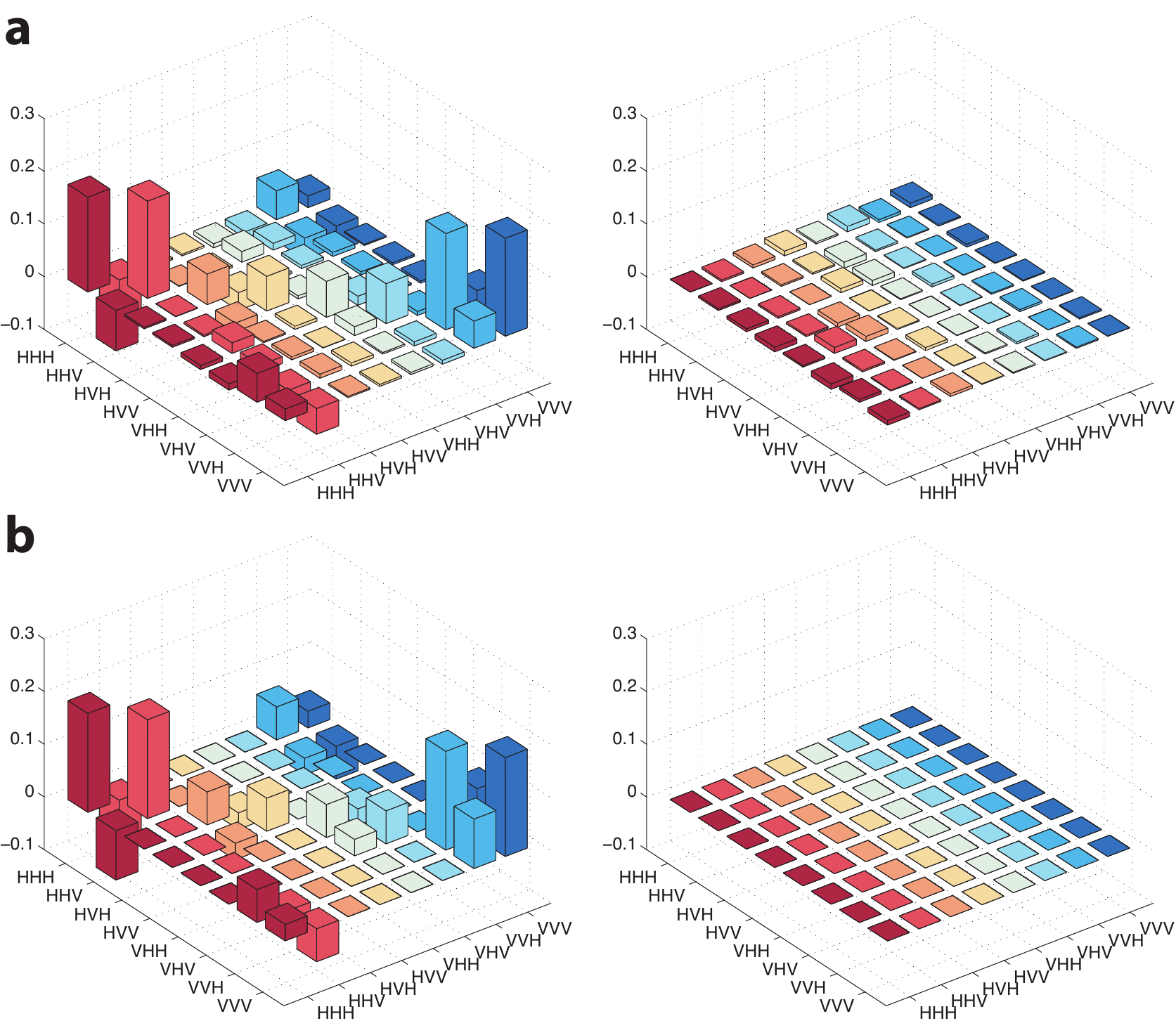}
\end{center}
\caption{Real (left) and imaginary (right) parts of the output state $\beta_{ABC}$.
(a) Experimental density matrix, obtained via three-qubit state tomography. (b) Ideal density matrix.
The fidelity between (a) and (b) is $98\%$, see text.}
 \label{FIG_Tomography}
\end{figure}

The circuit diagram in  Fig.~\ref{FIG_SETUP}a shows the conceptual implementation of the protocol, while the experimental setup is shown in Fig.~\ref{FIG_SETUP}b. Using four single photons---one of which acts as a trigger with the other three as the qubits $A$, $B$, and $C$---we prepare the state $\alpha_{AB}$  by summing up the individual pure-state terms in Eq.~\eqref{INITIAL_AB}, with measurement acquisition times corresponding to the weights. A similar technique prepares the mixed state $\alpha_C$ 
and guarantees that the initial state $\alpha_{ABC}$ is separable. Systems $A$ and $C$ interact in a controlled-phase gate \cite{PhysRevLett.95.21}, before the output state $\beta_{ABC}$
is characterised through a three-qubit state tomography~\cite{James01}. The total integration time was $387$ h, during which we counted $\sim$$30000$ four-fold coincidence events. 
The reconstructed density matrix has a large overlap with the ideal state, quantified by a fidelity of $\mathcal{F}(\beta_{\textrm{exp}},\beta_{\textrm{ideal}})\equiv\textrm{Tr}((\beta_{\textrm{exp}}^{1/2}\beta_{\textrm{ideal}}\beta_{\textrm{exp}}^{1/2})^{1/2})^2{=}0.98$,  and is shown in Fig.~\ref{FIG_Tomography}. To estimate the uncertainty, we perform a Monte Carlo analysis based on 10000 Poissonian-distributed variations of the measured photon counts. The corresponding population of reconstructed density matrices is used to evaluate an average fidelity of $\mathcal{F}_{\textrm{est}}{=}0.967{\pm}0.007$, which is extremely close to the experimental value.

In order to experimentally study the resilience of the protocol against noise and to obtain an unambiguous signature for entanglement distribution with separable states,
we add increasing amounts of white noise to the initial state, thus obtaining $\tilde{\alpha}_{ABC}=(1-p){\alpha}_{ABC}+\frac{p}{8}\openone$. Previously, this method has been used to assess the generation of bound-entangled states~\cite{PhysRevLett.105.130301}. Theoretically, $\tilde{\alpha}_{ABC}$ allows entanglement distribution with separable carriers for all $p < \frac{1}{3}$.

\begin{figure}[!b]
\begin{center}
\includegraphics[scale=0.16]{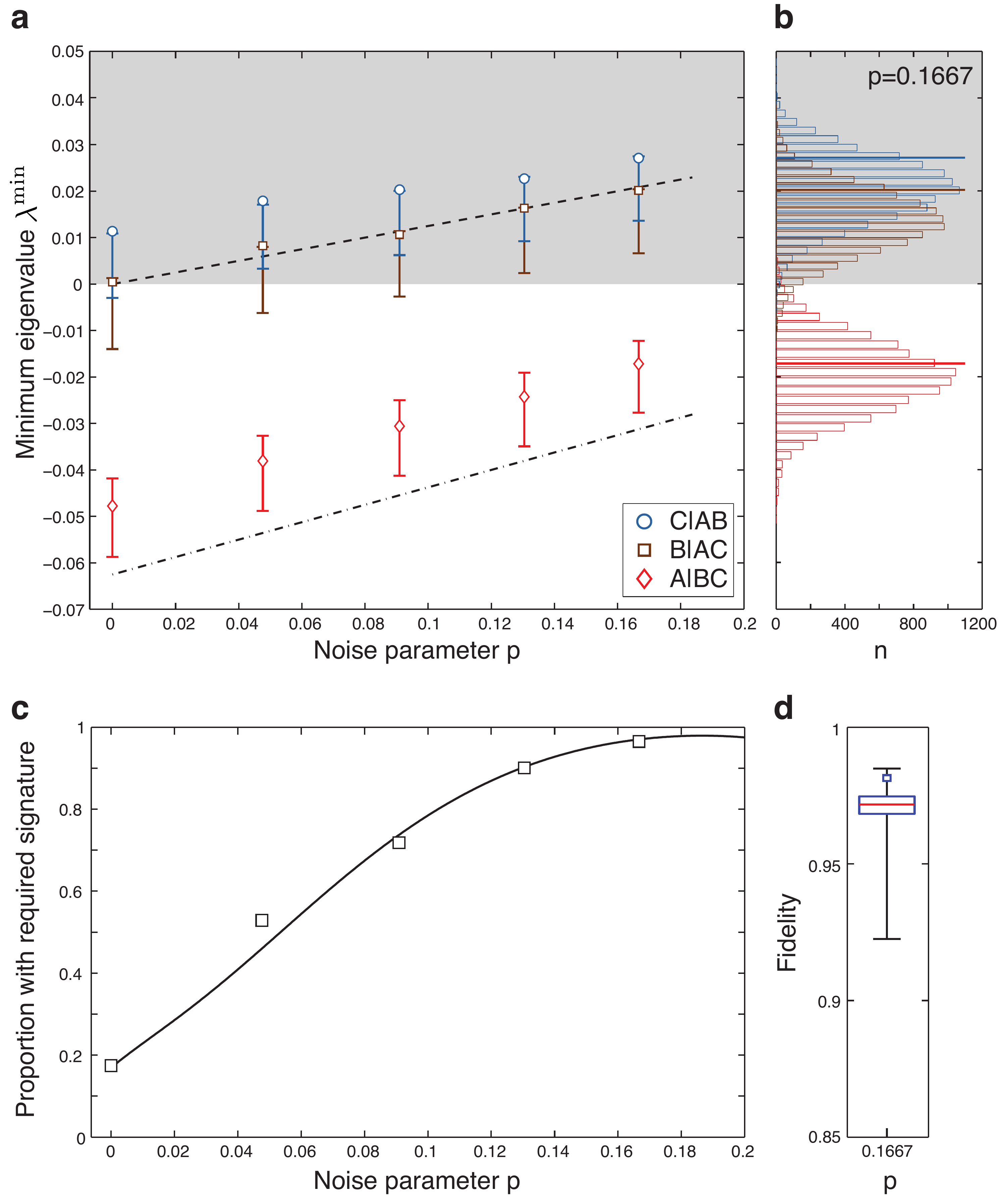}
\end{center}
\caption{(a) Minimum eigenvalue after partial transposition $\lambda^{\min}$ for each bipartition of $\beta_{ABC}$, against a white-noise admixture $p$. The dash-dotted (dashed) black lines show the theoretical values for infinite counts for the $A|BC$ ($C|AB$ and $B|AC$) bipartitions, respectively. Error bars represent 1 standard deviation of the distributions described in (b).
(b) $\lambda^{\min}$ for  $p=0.1667$, with experimental data (solid lines) and Monte Carlo distribution (histogram) based on a population of $10 000$ tomographic reconstructions with Poissonian variation of the measured counts.
(c) Proportion of the Monte Carlo population for which \emph{only} bipartition $\beta_{A|BC}$ has $\lambda^{\min}<0$. The solid line is a guide to the eye constructed by ideally adding white noise to the $p=0$ experimental state.  
(d) Box-and-whisker plot representing the fidelity distribution of the theoretical state with the Monte Carlo population for $p=0.1667$; the whiskers indicate maximum and minimum values. The data point represents the fidelity of the experimentally obtained state with the ideal one.
See Appendix~\ref{APP_STAT} for a discussion of statistical effects of limited photon counts in our experiment.}
 \label{FIG_EValues}
\end{figure}

In Fig.~\ref{FIG_EValues}a we plot $\lambda^{\min}$ for all bipartitions of the measured states as a function of added white noise. For $p=0$ only bipartition $A|BC$ is entangled, indicating a successful demonstration of the protocol. However, as shown in Fig.~\ref{FIG_EValues}c, only 17.4\% of the Monte Carlo population has the required success signature. This proportion rises rapidly with the addition of small amounts of white noise: 96.5\% of the population successfully demonstrates the protocol for $p = 0.1667$. 

The measured negativity with the maximum added noise ($p = 0.1667$) is $\mathcal{N}_{A|BC}^{\textrm{exp}}{=}{}0.0172$ with $\mathcal{N}_{B|AC}^{\textrm{exp}} = \mathcal{N}_{C|AB}^{\textrm{exp}} = 0$. In order to exclude the possibility that the controlled phase gate introduces bound entanglement~\cite{PhysRevLett.80.5239} that is distributed by system $C$, we provide an explicit decomposition of the experimental states in terms of convex sums of product states of the $C|AB$ bipartition in Appendix~\ref{APP_IDEAL} and~\ref{APP_EXP}.
We further show in Appendix~\ref{APP_DISCORD} that the experimentally distributed entanglement is strictly smaller than the communicated amount of quantum discord, confirming Eq.~(\ref{ENT_GAIN}).

A key question to address is the potential advantage of the protocol over other communication-based strategies for entanglement distribution. In this content, it is worth stressing that Alice and Bob will always do better by \emph{directly} sharing maximally entangled states, \emph{if} those are available~\cite{PhysRevLett.109.080503}. 
However, given noisy resources to start with---a reasonable assumption in any practical setting---we can identify regimes under which the distribution of entanglement via separable carriers is a winning strategy. 
In Appendix~\ref{APP_NOISE} we show that for depolarising and dephasing noise and starting from the paradigm resource embodied by Werner states, the protocol demonstrated here outperforms the direct sharing of entanglement. 
More specifically, we show that with such resources and under the action of the above quantum channels, the amount of distributed entanglement is higher using the protocol based on communication of separable states. Remarkably, in certain cases, only this scheme is able to distribute entanglement, thus demonstrating its practical value as an effective means to distribute entanglement across a network.

The fundamental insight that an important physical quantity can be increased in an experimental setting without transmitting that quantity reveals yet another counter-intuitive aspect of quantum mechanics. 
We demonstrated that distillable entanglement can indeed be distributed between remote parties who exchange only unentangled carriers. The success of our protocol is confirmed by the unambiguously entangled nature of the $A|BC$ bipartition and the separability of the other two. An equally interesting albeit weaker statement on entanglement distribution via bound-entangled states would be possible by having a $C|AB$ bipartition with positive partial transposition, but is not  separable. We have shown the robustness of the protocol to noise and the existence of experimentally relevant conditions under which distributing entanglement using a separable information carrier is indeed more advantageous than communicating entanglement between remote nodes of a network. 

\section*{Acknowledgements}

We would like to thank K. Kristinsson, P. Raynal, and K. \.Zyczkowski for discussions.
MZ and TP are supported by the National Research Foundation and Ministry of Education in Singapore, by the start-up grant of Nanyang Technological University, and NCN Grant No. 2012/05/E/ST2/02352.
MP thanks the UK EPSRC for financial support through a Career Acceleration Fellowship and a grant under the ``New directions for EPSRC research leaders" initiative; 
AF and MA for support from Australian Research Council Discovery Early Career Awards No. DE130100240 and No. DE120101899, respectively; AGW acknowledges support from a UQ Vice-Chancellor's Senior Research Fellowship.
This work was supported in part by the Centres for Engineered Quantum Systems (CE110001013) and for Quantum Computation and Communication Technology (CE110001027).

\emph{Note added}. Recently, we became aware of an independent demonstration of the phenomenon discussed here based on continuous-variable systems~\cite{arXiv:1303.1082,arXiv:1304:0504}.
These implementations make explicit use of the availability, in the Gaussian continuous-variable scenario, of necessary and sufficient criteria for the inseparability of tripartite mixed states.
The lack of similar tools in the discrete-variable case addressed in our work required the extra analysis reported in Appendix~\ref{APP_EXP}.
Moreover, differently from our protocol, the scheme realised in Ref.~\cite{arXiv:1304:0504} required the exchange of classical communication beside the communication of separable information carriers.

\appendix

\section{Separability in the ideal case}
\label{APP_IDEAL}

Here we provide a detailed analysis of the techniques used to reveal the separable nature, across the $C|AB$ bipartition, of the states that have been produced experimentally. We begin with the ideal theoretical case and later will apply some of the techniques  discussed here to the experimental density matrices.

Consider the initial state of system formed by qubits $A$ and $B$
\begin{eqnarray}
\alpha_{AB} & = & \frac{1}{4} \proj{HH} + \frac{1}{4} \proj{VV} \nonumber \\
&+ & \frac{1}{8} \proj{DD} + \frac{1}{8} \proj{AA} \nonumber \\
& + & \frac{1}{8} \proj{\carr\call} + \frac{1}{8} \proj{\call\carr},
\end{eqnarray}
where $\ket{H}$ ($\ket{V}$) denotes horizontal (vertical) polarisation state,
$\ket{D}$ ($\ket{A}$) denotes $45^\circ$ ($135^\circ$) linear-polarisation state,
and $\ket{\carr}$ ($\ket{\call}$) denotes right (left) circular polarisation state
\begin{eqnarray*}
\ket{D} = \frac{1}{\sqrt{2}}(\ket{H} + \ket{V}), & \quad & \ket{A} = \frac{1}{\sqrt{2}}(\ket{H} - \ket{V}), \\
\ket{\carr} = \frac{1}{\sqrt{2}}(\ket{H} + i \ket{V}), & \quad & \ket{\call} = \frac{1}{\sqrt{2}}(\ket{H} - i \ket{V}).
\end{eqnarray*}
These are the embodiment of the logical states $\{\ket{z_0},\ket{z_1}\}$, $\{\ket{x_0},\ket{x_1}\}$, and $\{\ket{y_0},\ket{y_1}\}$ introduced in the main manuscript.
The carrier qubit is initially in the state:
\begin{eqnarray}
\alpha_{C} = \frac{1}{4} \proj{D} + \frac{3}{4} \proj{A}.
\end{eqnarray}
After applying a controlled-phase gate on qubits $A$ and $C$, the overall system's state becomes
\begin{equation}
\beta_{ABC} = 
\left(
\begin{array}{cccccccc}
\frac{3}{16} & - \frac{3}{32} & . & . & . & . & \frac{1}{16} & \frac{1}{32} \\
- \frac{3}{32} & \frac{3}{16} & . & . & . & . & - \frac{1}{32} & - \frac{1}{16} \\
. & . & \frac{1}{16} & - \frac{1}{32} & . & . & . & . \\
. & . & - \frac{1}{32} & \frac{1}{16} & . & . & . & . \\
. & . & . & . & \frac{1}{16} & \frac{1}{32} & . & . \\
. & . & . & . & \frac{1}{32} & \frac{1}{16} & . & . \\
\frac{1}{16} & - \frac{1}{32} & . & . & . & . & \frac{3}{16} & \frac{3}{32} \\
\frac{1}{32} & - \frac{1}{16} & . & . & . & . & \frac{3}{32} & \frac{3}{16}
\end{array}
\right),
\end{equation}
where the density matrix is written in the standard basis and the dots represent zeros.
This state is entangled across the cut $A|BC$ as revealed by the partial transposition criterion:
the smallest eigenvalue of the corresponding partially transposed density matrix equals $- \frac{1}{16} = - 0.0625$.
The state is also separable across the $C|AB$ cut, as shown by the following explicit decomposition into product states for this bipartition
\begin{widetext}
\begin{eqnarray}
\beta_{ABC} & = & \frac{3}{16} \proj{HH} \otimes \proj{A} + \frac{3}{16} \proj{VV} \otimes \proj{D} \nonumber \\
& + & \frac{1}{8} \proj{\phi^+} \otimes \proj{H} + \frac{1}{8} \proj{\phi^-} \otimes \proj{V} \nonumber \\
& + & \frac{1}{16} \proj{HV} \otimes \proj{A} + \frac{1}{16} \proj{VH} \otimes \proj{D} \nonumber \\
& + & \frac{1}{16} \proj{\phi^{+i}} \otimes \proj{\call} + \frac{1}{16} \proj{\phi^{-i}} \otimes \proj{\carr} \nonumber \\
& + & \frac{1}{32} \proj{HV} \otimes \proj{\call} + \frac{1}{32} \proj{VH} \otimes \proj{\call} \nonumber \\
& + & \frac{1}{32} \proj{\psi^+} \otimes \proj{\carr} + \frac{1}{32} \proj{\psi^-} \otimes \proj{\carr},
\label{IDEAL_SEP}
\end{eqnarray}
\end{widetext}
where $\{\ket{\psi^{\pm}}, \ket{\phi^{\pm}}\}$ represents the standard Bell basis and $\ket{\phi^{\pm i}} = \frac{1}{\sqrt{2}}(\ket{HH} \pm i \ket{VV})$.

\section{Separability of the experimental data}
\label{APP_EXP}

In the experiment, additionally to producing states close to the ideal state $\beta_{ABC}$, we also prepared the set of states with increasing amount of admixed white noise.
All of such states are entangled in the cut $A|BC$, as demonstrated in the main text  by the existence of a negative eigenvalue in the spectrum of the matrix obtained after partial transposition of $A$. Figure 4a also shows that the other two cuts are associated with states having positive eigenvalues after partial transpositions. As this criterion is not a necessary and sufficient one for three-qubit states, this does not exclude the possibility of having bound entanglement in the one of the cuts that is positive under partial transposition. 

Therefore, in order to exclude the possibility of performing entanglement distribution via bound entangled states, we explicitly show separability across the bipartition $C|AB$ by constructing a separable decomposition of the corresponding states, in analogy with Eq. (\ref{IDEAL_SEP}).
To this end, we use the following algorithm:
\begin{enumerate}
\item We generate a set of random product states for the chosen cut and use them to complement the set of product state vectors that enter the decomposition given in Eq.~(\ref{IDEAL_SEP}).
\item We write a separable state as
\begin{equation}
\rho_{AB|C} = \sum_j p_j \proj{\pi_j},
\end{equation}
where $j$ is a label for the chosen product states $\ket{\pi_j}$ discussed at step 1, and $p_j$ are the associated probabilities of occurrence ($\sum_j p_j=1$).
\item We equate this expression with the experimental state and numerically solve for $p_j$'s.
\end{enumerate}
Only about $3000$ product states are sufficient to find explicitly the  separable decompositions of all experimental density matrices reported in the main manuscript and thus wash out any possibility for bound entanglement. The protocol is thus faithfully based on the use of separable states.

\section{Statistical effects of limited photon counting}
\label{APP_STAT}

\begin{figure}
\begin{center}
\includegraphics[scale=0.28]{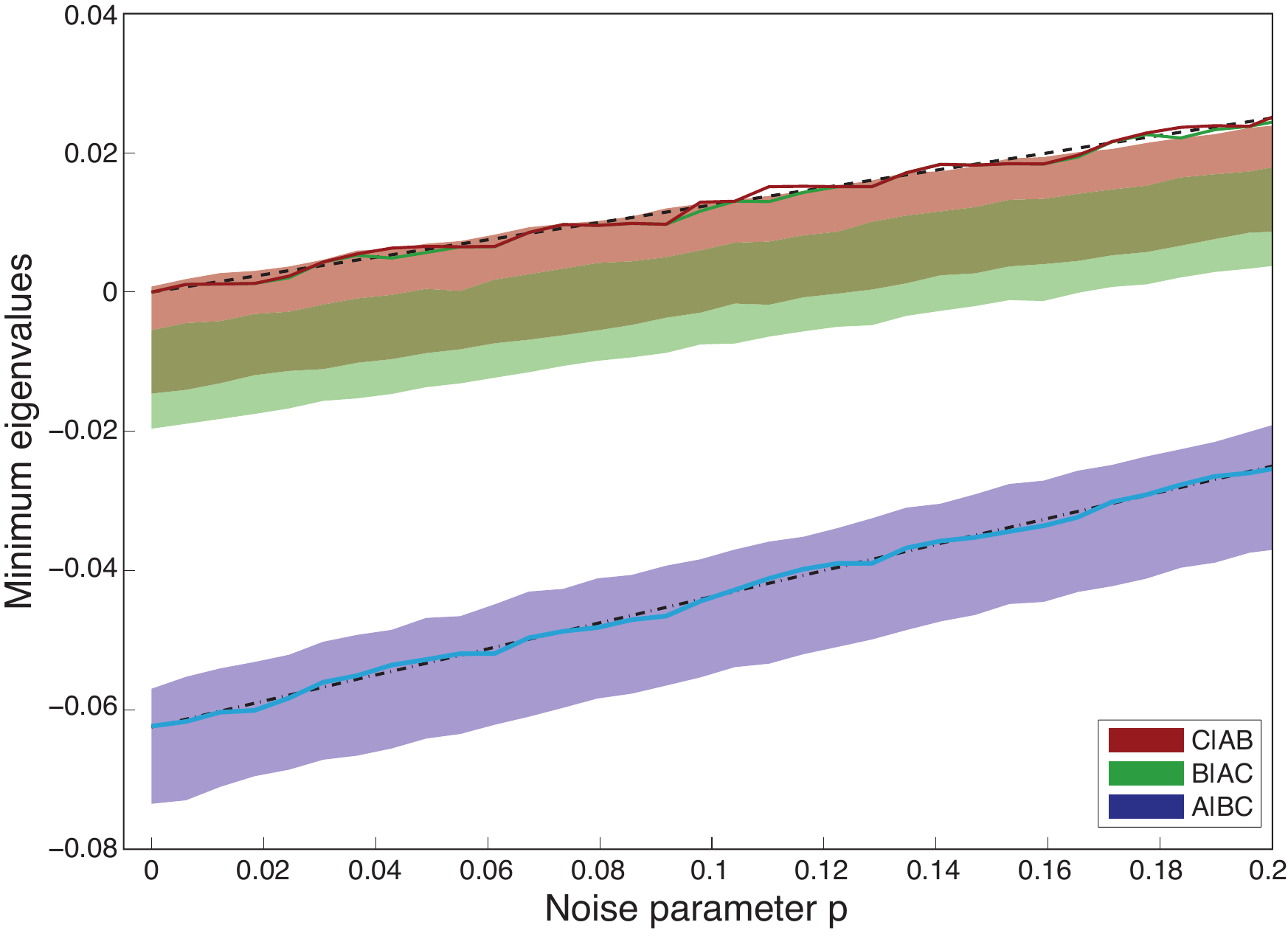}
\end{center}
\caption{Minimum eigenvalues after partial transposition on the first subsystem vs noise parameter $p$ for different bipartitions of the density matrix. The dashed (dash-dotted) black lines show the theoretical values for the unentangled, $C|AB$ and $B|AC$, (entangled $A|BC$) bipartitions respectively, assuming an infinite number of counts. The solid blue, green and red lines show the eigenvalues obtained from reconstructed density matrices starting with the ideal theory states but simulated with equivalent finite count statistics to those in the experiment. The shaded regions show the one-standard-deviation range obtained from adding Poissonian counting statistics to the ideal states. We observe a similar offset from theoretical states simulated with infinite counts, as in the experimental data (see Fig.4 in main text), including the splitting of the eigenvalues for the unentangled cuts. Each data point is calculated from $1000$ simulated density matrices of $\alpha_{ABC}$ at $50$ separate noise values.}
 \label{FIG_LimitedCounts}
\end{figure}

Our error analysis showed that our data points, Fig. 4 in the main text, lie slightly outside the most likely range obtained via the standard method for single-photon experiments: maximum-likelihood estimation of states with Poisson variation of the measured photon counts. In addition, we observe that the minimum eigenvalues for the unentangled bipartitions split from the theoretical value, with $B|AC$ having a more negative bias than $C|AB$. 

In Fig.~\ref{FIG_LimitedCounts} we present results of a similar analysis starting with ideal states and an ideal gate operation. These numerical simulations show the same effects, highlighting that they are not an artefact of non-ideal state preparation or gate imperfections, but solely due to statistics based on finite photon counts.

\section{Discord as bound on distributed entanglement}
\label{APP_DISCORD}

\begin{figure}
\begin{center}
\includegraphics[scale=0.45]{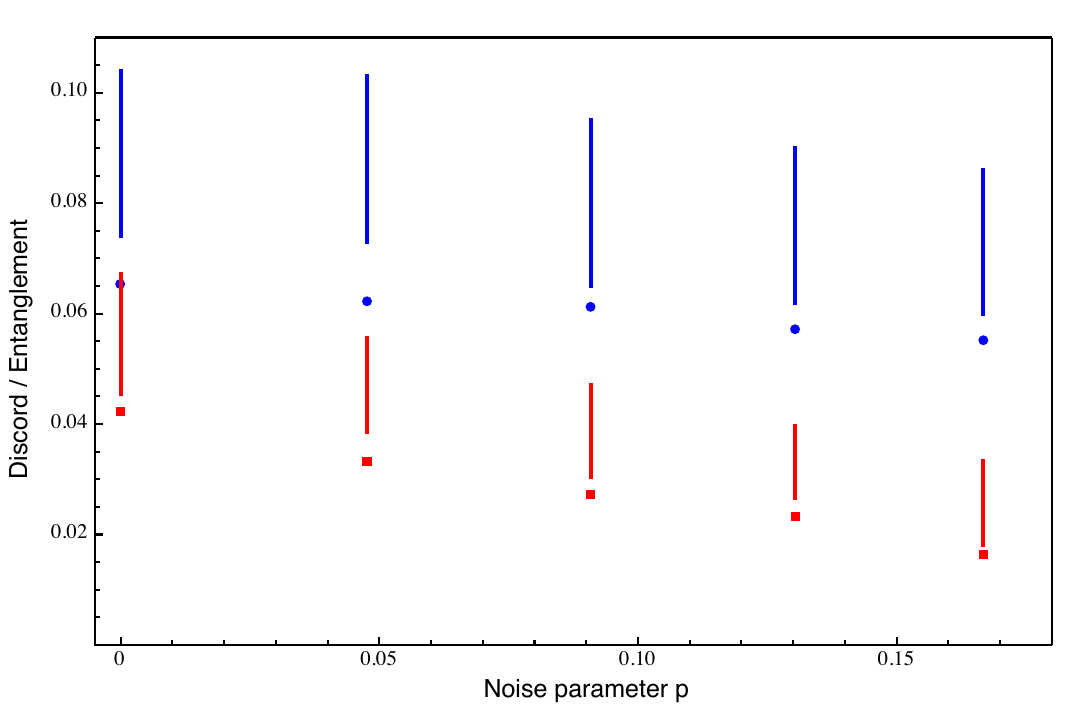}
\end{center}
\caption{Quantum discord bounds the amount of distributed entanglement. 
The dots give the value of the relative entropy of discord $\mathcal{D}_{\mathrm{comm}} = \mathcal{D}_{AB|C}$ for the reconstructed density matrices. 
The vertical blue lines present error bars for the discord as established by calculations of the discord for the hundred matrices obtained by adding Poissonian noise.
The squares give the upper bound on the final relative entropy of entanglement, $\mathcal{E}_{\mathrm{final}} = \mathcal{E}_{A:CB}$, and the red vertical lines represent error bars for the upper bound on the entanglement.
The offset of the statistical means with respect to the values of both the relative entropy of discord and the entanglement bound calculated from the experimentally reconstructed states can be explained analogously to what has been done in the previous Section. As the entanglement in the initial states of our system is null ($\mathcal{E}_{\mathrm{initial}} = \mathcal{E}_{AC:B} = 0$), this plot validates Eq. (2) of the main text and shows that discord is not the only relevant quantity for entanglement distribution because it does not provide a tight bound to the amount of entanglement generated by the protocol that we have implemented.}
\label{FIG_DE}
\end{figure}

In this Section we use the experimental data to validate the bound stated in Eq.~(2) of the main text.
The results are presented in Fig.~\ref{FIG_DE} and show that the discord is not the only relevant resource for entanglement distribution. In fact, there is a statistically significant gap between 
the amount of distributed entanglement and the communicated discord.

The relative entropy of discord $\mathcal{D}_{AB|C}$ is found numerically by minimising the entropic cost of a projective measurement on $C$. That is
\begin{equation}
\mathcal{D}_{AB|C}(\beta) = \min_{\Pi_C} [S(\Pi_C(\beta))] - S(\beta),
\end{equation}
where $S$ is the von Neumann entropy and $\Pi_C$ denotes a rank-one projective measurement on qubit $C$,
i.e. $\Pi_C(\beta) = \Pi_0 \beta \Pi_0 + \Pi_1 \beta \Pi_1$, with $\Pi_0$ and $\Pi_1$ the orthogonal projectors acting on qubit $C$, and $\beta$ is the reconstructed density matrix.

To find the upper bound on relative entropy of entanglement we use its definition
\begin{equation}
\mathcal{E}_{A:CB}(\beta) = \min_{\rho_{A:CB}} [S(\beta || \rho_{A:CB})],
\end{equation}
where the minimum is taken over all separable states $\rho_{A:CB}$ in the splitting $A|CB$, and the relative entropy is defined as $S(\beta || \rho_{A:CB}) = -\Tr(\beta \log {\rho_{A:CB}}) - S(\beta)$.
We used a mixture of sampling over the separable states together with numerical optimisations within classes of separable states to obtain small values of the relative entropy which give the upper bound on the relative entropy of entanglement.

\section{Robustness to noise}
\label{APP_NOISE}

In this Section we present a detailed analysis of the resilience of the protocol for the distribution of entanglement through separable states (EDSS) to the effects of relevant quantum channels acting on the communicated systems.  
This serves as a meaningful ground for a first assessment of any potential advantage of EDSS schemes over the most natural way of distributing entanglement, i.e. sending to Bob one particle out of a system of two that have been prepared in an entangled state. We denote this natural way as a \emph{direct} protocol. 
After sending $B$ to Bob, Alice can try to increase the amount of shared entanglement by  applying a controlled-phase gate locally on $A$ and $C$, and then sending the latter to Bob.
This protocol shall be referred to as an \emph{indirect} protocol when we do not distinguish whether transmitted particle $C$ is separable or entangled with the remaining particles.
Although we use the terms EDSS here, in principle (and very unlikely) the distribution could also be via bound entangled states as our calculations below are based on the partial transposition criterion, which is not necessary and sufficient for separability of three-qubit states.
The results show that Alice and Bob should choose to implement the EDSS protocol if they do not have access to relatively highly pure initial states.

As the initial state we consider the two-qubit Werner state
\begin{equation}
\rho_W = (1-p) \ket{\phi_-}\bra{\phi_-} + p \openone/4,
\label{WERNER}
\end{equation}
and system $C$ initialised in $\alpha_c=(1/2)(\openone+c_x\sigma_x)$ with $c_x=\frac{1}{3}$.
We have chosen to study the Werner states mainly because of their simplicity and clarity of exposition.
It will already become clear that EDSS is a useful protocol in some regimes.
Nevertheless, we would like to note that different initial states may exist, as well as different noises, that could show even more striking effectiveness of EDSS protocols.
Another argument to consider the initial Werner states relies on the fact that they are maximally entangled mixed states~\cite{MEMS}. 
More precisely, if mixedness is measured by the linear entropy and entanglement by the negativity, Werner states contain the highest amount of entanglement for a given mixedness.
Therefore all states with linear entropy corresponding to separable Werner states are also separable.
Clearly, they are all useless for direct entanglement distribution but we will now show that some entanglement can still be distributed via EDSS.

\begin{figure}
\begin{center}
\includegraphics[scale=0.38]{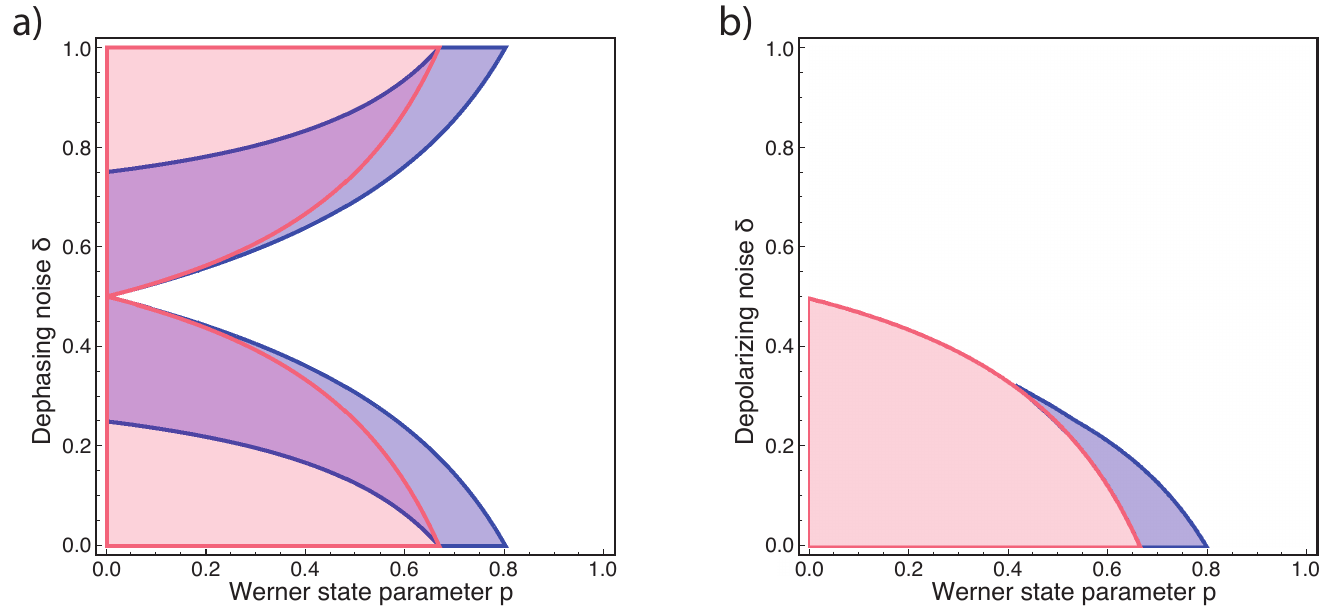}
\end{center}
\caption{Comparison between direct entanglement distribution protocol and the distribution via separable states for (a) dephasing and (b) depolarising channel.
In the protocol of direct distribution Alice prepares a Werner state (\ref{WERNER}) parameterised by $p$ and sends one qubit to Bob via a channel parameterised by $\delta$.
Entanglement is successfully distributed in the pink region.
Running the protocol of the main text beginning with the state distributed directly, entanglement is established via separable states within the blue area.}
\label{FIG_AREA}
\end{figure}

\begin{figure}[!b]
\begin{center}
\includegraphics[scale=0.35]{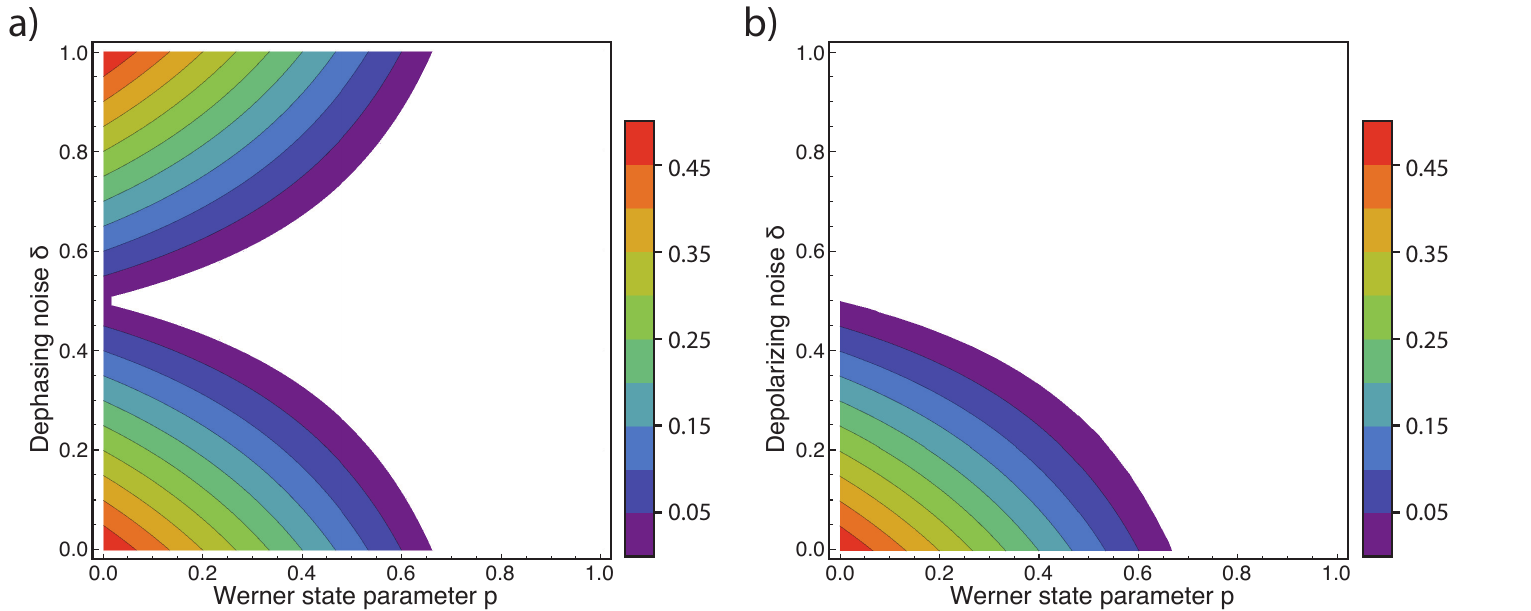}
\end{center}
\caption{Distributed entanglement, as measured by the negativity $A|BC$, obtained via the direct protocol in (a) dephasing and (b) depolarising channel.}
\label{FIG_DIRECT}
\end{figure}

\begin{figure}
\begin{center}
\includegraphics[scale=0.35]{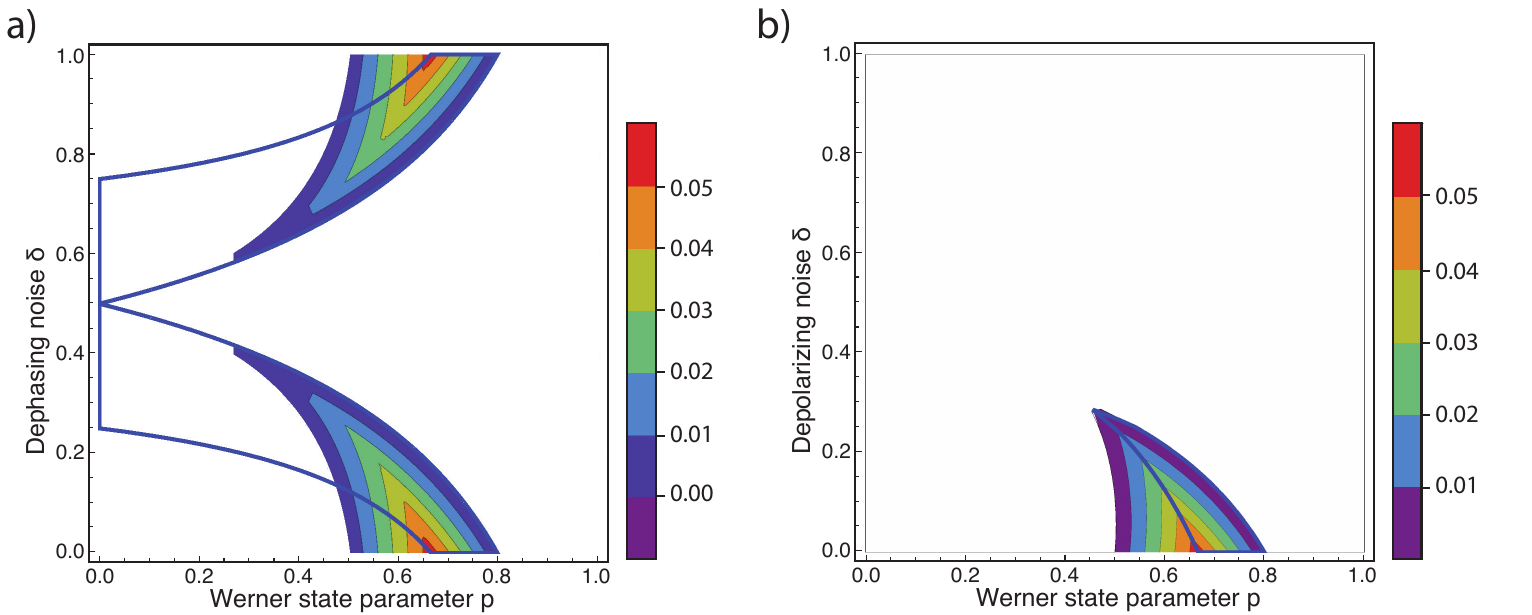}
\end{center}
\caption{Entanglement gain, as measured by the difference in the negativity $A|BC$, between the final state after the indirect protocol and after the direct one, (a) for the dephasing channel and (b) for the depolarising channel.}
\label{FIG_GAIN}
\end{figure}

We begin with the dephasing channel acting on the communicated system.
The channel is described by the set of Kraus operators $\{\sqrt{1-\delta} \openone,\sqrt{\delta} \sigma_z\}$,
where $\delta\in[0,1]$ is the strength of the environmental action, and we focus on $\delta \ge \frac{1}{2}$ as in this case the effects of the channel are symmetric about this value.
As Fig.~\ref{FIG_AREA} shows, there exists a range of the parameters $p$ and $\delta$ for which EDSS is the only working protocol for distributing entanglement.
As a measure of robustness of an entanglement distributing protocol against noise 
we compute the area  $R$ of the plot in Fig.~\ref{FIG_AREA} for which the distribution is successful.
The larger the value of $R$, the more robust we consider the corresponding protocol as more initial states and noises lead to final entanglement.
For the direct protocol with sending a half of the Werner state we find $R_{\textrm{direct}} = 0.45$
and for the distribution via separable states $R_{\textrm{edss}} = 0.30$.
Therefore, not only EDSS allows the new parameter regime for successful entanglement distribution,
but the protocol is also quite robust with $R_{\textrm{edss}}$ being almost $70\%$ of $R_{\textrm{direct}}$.
Naturally, one is also interested in the amount of distributed entanglement.
In Fig.~\ref{FIG_DIRECT} we plot the negativity obtained in the direct protocol, while Fig.~\ref{FIG_GAIN} shows the difference between the negativity in the partition $A|BC$ after the indirect protocol and after the direct one. 
Note the increment in the entanglement via the EDSS protocol for many values of $p$ and $\delta$, in some cases being significant.

Next, we study the effect of depolarising channel on the same scenario.
The depolarising channel is described by the set of Kraus operators $\{\sqrt{1-\delta}\openone,\sqrt{\frac{\delta}{3}}\sigma_x,\sqrt{\frac{\delta}{3}}\sigma_y,\sqrt{\frac{\delta}{3}}\sigma_z\}$. 
The robustness of the direct protocol equals $R_{\textrm{direct}} = 0.23$ whereas EDSS has $R_{\textrm{edss}} = 0.026$, being only $10\%$ of $R_{\textrm{direct}}$ (see Fig. \ref{FIG_AREA}).
However, we notice that the range in which EDSS distributes entanglement is completely disjointed from the direct one.
Hence, EDSS strictly enlarges possible combinations of $p$ and $\delta$ that can be used to obtain entanglement. 
Figs.~\ref{FIG_DIRECT} and~\ref{FIG_GAIN} show that a significant entanglement gain can be achieved by using the EDSS protocol.

\end{document}